\documentclass[12pt]{article}

\input epsf
\newcount\figno
\def\fig#1#2#3{
\par\begingroup\parindent=0pt\leftskip=1cm\rightskip=1cm\parindent=0pt
\baselineskip=11pt
\global\advance\figno by 1
\epsfxsize=#3
\centerline{\epsfbox{#2}}
\vskip 12pt
{\bf Figure \the\figno:} #1\par
\endgroup\par
}
\def\figlabel#1{\xdef#1{\the\figno
\mbox{ }}}
\def\encadremath#1{\vbox{\hrule\hbox{\vrule\kern8pt\vbox{\kern8pt
\hbox{$\displaystyle #1$}\kern8pt}
\kern8pt\vrule}\hrule}}

\def\href#1#2{#2}
\def\beq{\begin{equation}}
\def\eeq{\end{equation}}

\def\drawbox#1#2{\hrule height#2pt 
        \hbox{\vrule width#2pt height#1pt \kern#1pt 
              \vrule width#2pt}
              \hrule height#2pt}

\def\Asym#1#2{\vcenter{\vbox{\drawbox{#1}{#2}
              \kern-#2pt       
              \drawbox{#1}{#2}}}}

\begin{document}

\begin{titlepage}

\begin{center}
\today
\hfill HUB-EP 98/22\\
\hfill hep-th/9803232
\vspace{2cm}

{\Large\bf Equivalence of Geometric Engineering and Hanany-Witten via
Fractional Branes}

\vspace{2cm}
{\large 
Andreas Karch, \,  Dieter L\"ust 
\, and \,\,
Douglas J. Smith}

\vspace{1cm}

{\it
Institut f\"ur Physik,
Humboldt Universit\"at Berlin\\
Invalidenstr. 110\\
10115 Berlin, Germany}

\vspace{.5cm}
{\tt karch@qft1.physik.hu-berlin.de \\
     luest@qft1.physik.hu-berlin.de \\
     smith@qft2.physik.hu-berlin.de}

\end{center}

\vspace{1.5cm}

\begin{abstract}

We present an explicit relation between the Hanany-Witten and Geometric
Engineering approaches of realising gauge theories in string theory. The
last piece in the puzzle is a T-duality relating arbitrary Hanany-Witten
setups and fractional branes.

\end{abstract}

\end{titlepage}

\section{Introduction}

During recent years it has become clear how supersymmetric quantum field
theories can be derived from string theory in the limit where
gravitational effects decouple from the gauge interactions, i.e. in
the limit $M_{\rm Planck}\rightarrow\infty$.
In this context there exist two competing scenarios, namely first
the `Geometric Engineering' approach
\cite{Klemm,EngineerII,KMV} and second the Hanany-Witten
\cite{HW} brane set up.

Within Geometric Engineering one considers
type IIA/B string compactification on a complex Calabi-Yau space.
For example type IIA/B compactification on a Calabi-Yau three-fold
leads to ${\cal N}=2$ supersymmetric field theories in four dimensions.
Perturbative as well non-perturbative effects
in ${\cal N}=2$ field theory can be explained by the
geometry of the underlying Calabi-Yau spaces.
Non-Abelian gauge fields together with massless matter
fields arise due to the wrapping of
D-branes around internal Calabi-Yau cycles;
as soon
as these cycles  shrink to zero size the corresponding fields become massless.
Since one assumes that the Calabi-Yau 3-fold is a $K3$-fibration over
a sphere,
the resulting gauge groups are immediately determined
by the singular $K3$-fibers, which are classified according
to the simple ADE algebras.
In the field theory limit all the interesting information is encoded
in the local geometry around the singularity, which means that
the field theory limit is performed by
replacing  the compact $K3$ fiber by a non-compact ALE space that
allows for the same ADE singularity classification as $K3$.
The non-perturbative effective action a la Seiberg and Witten is given
in terms of the rational instanton numbers of the Calabi-Yau space,
the  type IIA superstring is compactified on. It is very beautiful
that these rational instanton  numbers can be explicitly computed
by the mirror map to the corresponding type IIB model. Restricting
oneself only to the  properties of the
local Seiberg-Witten geometry around the singular cycles, the
full non-perturbative solution of the string derived
${\cal N}=2$ gauge theory is provided
by the local mirror map acting on the ALE space fibered over a sphere.

On the other hand, in the Hanany-Witten approach supersymmetric field theories
arise as the world-volume theories of certain brane configurations, which
fill the non-compact, say four-dimensional, spacetime.
As soon as parallel D-branes approach each other, the corresponding
non-Abelian gauge
bosons become light. In this framework, the solution of the non-perturbative
dynamics is provided by embedding the type IIA brane configurations
into M-theory, i.e. lifting all branes to their 11-dimensional parent
configurations. Then all singularities are smeared out, and the Seiberg-Witten
curves are rediscovered in a very nice way.

Since M-theory provides a unified description of all superstring theories
and their interactions, one expects on general grounds that the
Geometric Engineering approach and the Hanany-Witten brane set up are
equivalent to each other and are connected by some duality symmetries. In fact,
a vanishing cycle ($A_1$ singularity)
is  T-dual to two approaching branes, as was already discussed
in \cite{dualNS}. In this paper
we will explicitly show how this unification between the two ways
to derive ${\cal N}=2$ supersymmetric field theories from strings is realized.
Concretely, we will demonstrate that the Hanany-Witten brane configurations
are connected to fractional D-branes (which arise when considering D-branes
on orbifolds \cite{DouglasMoore, Enhanced})
via T-duality. Starting from the Geometric Engineering side, the IIA string
on a Calabi-Yau space is also related to the fractional brane configurations
by  combined S- and T-duality transformations. Therefore, the
fractional branes are half way between the Calabi-Yau compactifications
with branes wrapped around the internal cycles, and the world-volume
gauge theories with spacetime filling branes.

In Section 2 we will review the various approaches to the subject. In Section
3 we will propose the duality relation which allows us to connect
the various approaches and treat them in a unified way. Even
though this construction is general, we will focus on the
approaches yielding ${\cal N}=2$ supersymmetry in 4 dimensions. The connection
will be explained in Section 4. In Section 5 we discuss generalizations
to the ${\cal N}=1$ case and in Section 6 we conclude.

\section{The pieces of the puzzle}

There have been various approaches to embedding field theory in string
theory. The most prominent ones are the Geometric Engineering approach,
the Hanany-Witten approach and using Branes as Probes. We will
briefly review certain aspects of these 3 approaches before we move on
to show that they all fit nicely into a unified picture.

\subsection{Geometric Engineering}

\subsubsection{The setup: IIA on K3 fibered CY}

Four-dimensional ${\cal N}=2$ supersymmetric string vacua are obtained by
compactifying the heterotic string on $K3\times T^2$ or via string-string
duality \cite{KV,FHSV} by compactifying
the type IIA/B superstring
on a Calabi-Yau three-fold (for a recent review see \cite{DL}).
The ${\cal N}=2$ heterotic/type II string duality can be explained,
at least in a
heuristic way, from the six-dimensional string-string duality between the
heterotic string on $T^4$ and the type IIA string on
$K3$. Using an adiabatic argument and applying the
six-dimensional  duality fiber wise,
the resulting space on the type IIA side is
a $K3$ fibered Calabi-Yau space $X^3$ with base $P^1$ \cite{KLM,KLM2};
the various ways
to perform this $K3$-fibration
results in the variety of different type II ${\cal N}=2$
vacua.
Later on we will allow for more general two-dimensional bases, namely
intersecting $P^1$'s.

The massless spectrum is determined
by the cohomology of the three-fold, namely by the two Hodge numbers
$h^{(1,1)}$ and $h^{(2,1)}$.
As it is known, the type IIA/B
compactifications are related by mirror symmetry, which means
that  the type IIA superstring on the Calabi-Yau space $X^3$
is equivalently described by compactifying the type IIB on the mirror
manifold $\tilde X^3$ which has exchanged Hodge numbers: $\tilde h^{(1,1)}=
h^{(2,1)}$, $\tilde h^{(2,1)}=h^{(1,1)}$.
To be specific, the massless spectrum of the type IIA superstring contains
$h^{(1,1)}+1$ $U(1)$ ${\cal N}=2$ vector multiplets.
Their $h^{(1,1)}$ complex scalar moduli in the NS-NS
sector correspond to the deformations of the K\"ahler form $J$ of
$X^3$ plus the internal
$B_{MN}$ fields; the $h^{(1,1)}$ $U(1)$ R-R vectors  originate from
the ten-dimensional 3-form gauge potential $A_{MNP}$ with two indices
in the internal space. So the Abelian gauge symmetry including
the graviphoton is given by
$U(1)^{h^{(1,1)}+1}$. Finally there are $h^{(2,1)}+1$ massless ${\cal N}=2$
hypermultiplets. $h^{(2,1)}$ of them correspond to the complex structure
deformations of $X^3$.
The additional
hypermultiplet contains together with the
NS-NS axion field $a$ the four-dimensional dilaton $\phi^4_{IIA}$
plus two  more R-R scalar fields.
Since the dilaton belongs to a hypermultiplet in the
four-dimensional type II vacua it follows that the type II vector couplings
(gauge couplings plus moduli space metric)
do not depend on the type II dilaton and are purely classical in string
perturbation theory. Moreover, in type IIB compactifications, the vector
multiplets correspond to complex structure moduli, and hence their interactions
are not even affected by world-sheet instantons. This fact allows to calculate
exactly all IIB vector couplings in terms of pure geometry.

To discuss the non-Abelian gauge symmetries  in the type IIA string,
we have to find the charged BPS states which couple to Abelian R-R
gauge bosons \cite{HullTown}.
These cannot be found within the perturbative
type II spectrum, but they are
given by the non-perturbative  D2 branes which can be wrapped
around the 22 homology two-cycles of $K3$.
The masses of the BPS particles are proportional to the area of the $K3$
two-cycles. Therefore  massless non-Abelian gauge bosons arise at those
points in the $K3$ moduli space where the $K3$ degenerates, and
the two-cycles shrink to zero sizes. However not all $K3$ cycles can lead
to massless gauge bosons since we have to take into account the global
structure due to the $K3$ fibration \cite{Asp}.
This means that not the whole intersection lattice
$\Lambda$ of $K3$ can lead to
a non-Abelian gauge group enhancement, but only the monodromy invariant part
of $\Lambda$. This is the so-called Picard lattice $\Lambda_\rho$,
whose rank is given by the Picard number $\rho$.
Moreover for the gauge bosons to be massless, the size of the base $P^1$,
over which the $K3$ is fibered, must be infinite. This limit corresponds on the
dual heterotic side to the perturbative limit with zero string coupling
constant. For finite size of $P^1$ the gauge bosons will be in general massive,
and instead one deals with massless hypermultiplets at some points
in the vector moduli space. This situation  precisely reflects
the non-perturbative ${\cal N}=2$ dynamics described by Seiberg and Witten.
Non-perturbatively, the classical singularities will split, and the quantum
moduli space contains special loci of massless BPS hypermultiplets. Therefore,
in
the type IIA string, the whole Seiberg Witten solution is described entirely
in geometric terms.

In summary, concerning the gauge group,  it arises
due to the ADE singularities of $K3$. In particular, if one deals with
a genus $g$ curve with ADE singularity one expects to have ADE gauge
symmetry with $g$ adjoint hyper multiplets \cite{BVS,KMP,KM}.
Hence the gauge theory
is asymptotically free for $g\leq 1$; for $g=1$ one obtains the ${\cal N}=4$
spectrum.
Other matter representations are obtained from different kind of singularities.
In particular if there is an ADE singularity over a surface and this
singularity is enhanced to a higher one at some points along the
surface, then there will be some matter localized at those points.
For example, if the $K3$ fiber possesses an $A_{n-1}$ singularity
which is enhanced to $A_n$ at $k$ points on the surface, we get an
$SU(n)\times U(1)$ gauge group with $k$ hyper multiplets in the
fundamental of $SU(n)$ and charged under the $U(1)$ \cite{BKKM}.
Other examples with various matter representations have been discussed in
\cite{BKKM}.

\subsubsection{Decoupling Gravity and ALE spaces}

Now we want to decouple gravitational and stringy effects to keep only
all  effects due to the gauge dynamics.
So we take the limit $\alpha'\rightarrow 0$ for
describing the low energy physics. In this limit only the local
neighborhood of the singularity is relevant. This will be independent
of
the details of the chosen Calabi-Yau containing the singularity in this limit.
This means that the local geometry is described by a non-compact ALE
space
\cite{Klemm}. This ALE
space singles out the relevant shrinking 2-cycles
around which the D2 branes are wrapped. Information
from more distant parts of $K3$, like
effects from 2-branes wrapping other 2-cycles, are suppressed by
powers of $\alpha'$.

As an illustrative example \cite{EngineerII} consider the geometric engineering 
of a pure $SU(2)$ gauge symmetry which related to an $A_1$ singularity
in $K3$. Locally, one needs a vanishing 2-sphere $P^1$,
around which the D-branes, being the $W^\pm$ bosons, are wrapped.
This $P^1_f$ has
to be fibered over the base $P^1_b$ in order to have ${\cal N}=2$ supersymmetry
in four dimensions. The different ways to perform this fibration are encoded
by an integer $n$, and the corresponding fiber bundles are the
Hirzebruch surfaces $F_n$.
The mass (in string units)
of the $W^\pm$ bosons corresponds to the area of the fiber
$P^1_f$, whereas the area of the base $P^1_b$ is proportional to
$1/g^2$ ($g^2$ is the four-dimensional gauge coupling at the string scale).
Now let us perform the field theory limit which means that we send the
string scale to infinity. The running of the gauge coupling implies that
${g^2}$ should go to zero in this limit; thus the K\"ahler class
of the base $P^1_b$ must go to zero: $t_b\rightarrow \infty$. Second,
in the field theory limit the gauge boson masses should go to zero,
i.e. $t_f\rightarrow 0$. In fact these two limits are related by
the running coupling constant,
\begin{equation}
{1\over g^2}\sim\log {M_W\over\Lambda},
\end{equation}
and the local geometry is derived from the following double scaling
limit:
\begin{equation}
t_b\sim\log t_f\rightarrow\infty .\label{doublesc}
\end{equation}

Clearly, this picture can be easily generalized to engineer higher ADE
gauge group. In this case there is not only one shrinking $P_f$ in the
fiber, but several such that the fiber acquires a local ADE singularity.
For the close comparison with the Hanany-Witten set up, it is also
interesting how product gauge groups can be geometrically
engineered \cite{KMV}. For concreteness consider the group $SU(n)\times
SU(m)$ with a hypermultiplet in the bi-fundamental representation.
To realize this we have an $A_{n-1}$ singularity over $P^1_b$ and an
$A_{m-1}$ singularity over another  $P^1_b$. The two $P^1_b$'s intersect
at a point where the singularity jumps to $A_{m+n-1}$. This can
be seen in 6 dimensions as symmetry breaking of $SU(n+m)$
to $SU(n)\times SU(m)\times U(1)$ by the vevs of some scalars in the
Cartan subalgebra of $SU(n+m)$. It is straightforward to generalize
this procedure to an arbitrary product of $SU$ groups with matter in
bi-fundamentals. One associates to each gauge group a base $P^1_b$ over
which there is the corresponding $SU$ singularity;
to each pair of gauge groups connected by a bi-fundamental representation
one associates an intersection of the base $P^1_b$'s, where over
the intersection point the singularity is enhanced to $SU(n+m)$.

\subsubsection{The Solution: Local Mirror Symmetry}

So far our discussing was mainly limited to the classical aspects
of the gauge theory derived from Calabi-Yau compactifications.
As it is well known, the classical moduli space of ${\cal N}=2$ supersymmetric
Yang-Mills theories is corrected by quantum effects.
Consider the ${\cal N}=2$ prepotential $F(A)$ of the pure $SU(2)$ gauge theory,
spontaneously broken to $U(1)$ by the vev $a$ of the scalar component
of the Abelian ${\cal N}=2$ vector multiplet $A$ \cite{SW,SW2}:
\begin{equation}
F(A)={1\over 2} \tau_0A^2+{i\over \pi}A^2\log\biggl({A\over\Lambda}\biggr)^2
+{1\over 2\pi i}A^2\sum_{l=1}^\infty c_l\biggl({\Lambda\over A}\biggr)^{4l}.
\label{fieldprep}
\end{equation}
Here $\tau_0={1\over g^2}$ is the classical gauge coupling, the logarithmic
term describes the one-loop correction due to the running of
the gauge coupling and the last term collects all non-perturbative
contributions from the instantons with instanton numbers $c_l$.
The solution to the problem of computing these instanton numbers in field
theory was provided by Seiberg and Witten \cite{SW,SW2} by introducing
an auxiliary Riemann surface, the Seiberg-Witten curve, whose period
integrals define the period  $a$ and the dual period $a_D$ in terms of a
gauge invariant order parameter $u$.

As already mentioned, the corresponding
prepotential in type IIA Calabi-Yau compactifications,
which depends on
the K\"ahler-class moduli $t_A$ ($A=1,\dots ,N_V=h_{1,1}$), is purely classical
in string perturbation theory and receives only corrections due
to world-sheet instantons. It has the following structure \cite{Cand}:
\begin{eqnarray}
 {\cal F}^{\rm II} = -{1\over 6}C_{ABC}t_At_Bt_C
- \frac{\chi \zeta(3)}{2 (2 \pi^{3})}+
\frac{1}{(2 \pi)^3} \sum_{d_1,...,d_h}
n_{d_1,...,d_h} Li_3({\bf e} [i\sum_Ad_At_A]),
\label{ftype2}
\end{eqnarray}
where we work inside the K\"ahler cone ${\rm Re}t_A\geq 0$.
The polynomial part of the type-IIA
prepotential is given in terms of the classical intersection
numbers $C_{ABC}$ and the Euler number $\chi$,
whereas the coefficients $n_{d_1,...,d_h}$
of the exponential terms denote the rational instanton numbers of
genus 0 and multi degree $d_A$.
The rational instantons are related to non-perturbative
effects on the world-sheet, namely they
count the  embeddings of the genus
0 world-sheet into the Calabi-Yau space; their
contributions disappear in the large radius limit
$\alpha'/R\rightarrow 0$.\footnote{Several checks had been made
that the type II prepotential agrees with the prepotential of the
dual heterotic compactification on $K3\times T^2$ \cite{KV,checksprep,checksprep2,
checksprep3}.}

The nice way to compute the world-sheet instanton
numbers is to perform the mirror
map to the type IIB superstring compactified on the mirror Calabi-Yau
space. In this way
all the necessary non-perturbative information can
be obtained by purely geometric considerations
on the mirror Calabi-Yau, namely
by computing the corresponding Picard-Fuchs equations.
In the type IIB picture one is considering vanishing 3-cycles of
the mirror Calabi-Yau threefold (corresponding to conifold singularities).
The relevant BPS states correspond to type IIB D3 branes wrapped around
those 3-cycles.
With respect to the $K3$ fibration structure, the type IIB 3-branes
are wrapped around the vanishing 2-cycles of $K3$ leaving in this way
not massless particles, but non-critical anti-self-dual strings on
the base. The tension of this string depends on the size of the relevant
2-cycle.

In order to extract the field theory instanton numbers $c_l$ and also
the logarithmic piece in eq.(\ref{fieldprep}) from the string prepotential
eq.(\ref{ftype2}) one has to perform the double scaling limit
(\ref{doublesc}) which replaces the $K3$ by the local ALE space.
Therefore this procedure is called the local mirror map.
In fact, the local mirror map  simplifies the rederivation
of the Seiberg-Witten action enormously
since only the data in the
neighborhood of the ALE singularity is relevant. To understand this
we briefly discuss the local mirror map for the pure $SU(2)$ gauge
symmetry \cite{EngineerII}. As explained in the last section all the local
information we need is encoded in the fibration of $P^1_f$
over $P^1_b$. Therefore the prepotential is a function of the
moduli $t_f$ and $t_b$. $n_{d_b,d_f}$ denote the number of world-sheet
instantons wrapping $d_b$ times around the base $P^1_b$ and $d_f$ times
around the fiber $P^1_f$.
First, in the perturbative limit only the instanton  numbers
$n_{0,d_f}$ contribute, and, using the limit (\ref{doublesc}),
the string prepotential eq.(\ref{ftype2})
precisely leads to the logarithmic term in eq.(\ref{fieldprep}).
Second, and even more interesting, summing over all rational instantons
$n_{d_b,d_f}$ with $d_b\neq 0$ and performing again the limit
(\ref{doublesc}) the Seiberg-Witten instanton numbers $c_l$ should be
derived as explained in \cite{EngineerII}. In fact, one should get the same
result for all Hirzebruch surfaces $F_n$.

Having obtained the field theory instanton numbers from string theory
via the local mirror map,
it is highly suggestive that the field theory Seiberg-Witten curve
can be directly derived from the string Calabi-Yau geometry. How this works
was explained in \cite{Klemm}.
We briefly outline the main arguments for this beautiful
picture.
Consider first the Seiberg-Witten curve in ${\cal N}=2$
supersymmetric Yang-Mills
gauge theory based on some gauge group $G$. The Seiberg-Witten curve
can be considered as a fibration of the corresponding weight diagram
(points) over $P^1$.
In string theory, the 0-dimensional local fibers will be replaced
by appropriate ALE spaces with corresponding vanishing 2-cycles.
Indeed, the vanishing 2-cycles of the ALE space behave exactly
like the root vectors of the corresponding ADE groups.
This means that after
the field theory limit $\alpha'\rightarrow 0$,
the 3-branes wrapped around the type IIB Calabi-Yau
3-cycles are equivalent to the anti-self-dual strings, wrapped around
the Seiberg-Witten curve.
So in the type IIB picture, the role of the vanishing 1-cycles of
the Seiberg-Witten curve is played by the vanishing 3-cycles of
the Calabi-Yau space.
In this way, the Seiberg-Witten curve has not only an auxiliary
function, but gets a real physical interpretation in string theory.
This solution is sketched in the following diagram.

\vspace{1cm}
\fig{Solving in Geometric Engineering via Local Mirror Symmetry}
{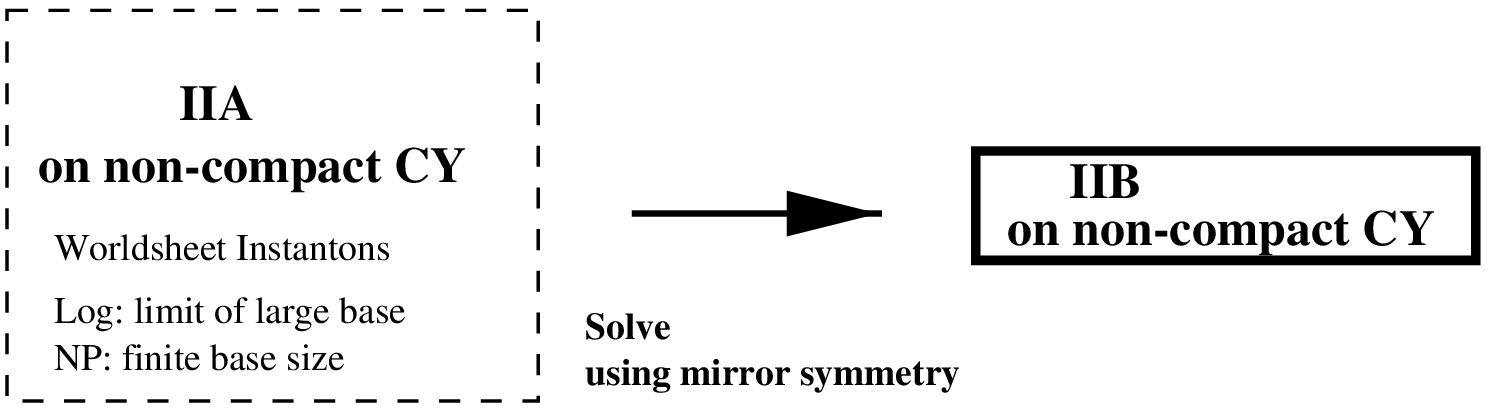}{10truecm}
\figlabel{\eng}

\subsection{The Hanany Witten setup}

Here we will review the method first used by Hanany and Witten to study 3
dimensional gauge theories \cite{HW} and since then 
generalised to other dimensions.
In this approach to studying gauge theories, D-branes are placed in flat space.
In this way all manifest properties of the gauge theory are encoded in the
positions and shapes of the branes. Here we will review the aspects of this
construction relevant to this paper. In particular we will consider the setup
appropriate for studying 4 dimensional ${\cal N}=2$ $SU(N)$ gauge theories.
\footnote{The gauge group is U(N) in low dimensions ($d < 4$) but in 4 or
more dimensions the U(1) factor is frozen out \cite{Witten}.}

\subsubsection{The setup}

To study ${\cal N}=2$ $SU(N)$ gauge theories in 4 dimensions Witten
\cite{Witten}
considered the
following  construction of branes in IIA string theory. $N$ parallel D4 branes
filling 01236 spacetime were bounded in the 6 direction by 2 NS5 branes
filling 012345 spacetime. This produced an effective 4 dimensional world-volume
for the D4 branes and it can easily be checked that ${\cal N}=2$
supersymmetry is
preserved in 4 dimensions for this setup. Matter can be added to the gauge
theory by adding semi-infinite D4 branes filling 01236 and ending ``outside''
the NS5 branes. Another method to produce matter is to add D6 branes filling
0123789 between the 2 NS5 branes in the 6 direction. These two constructions
are in fact related since moving a D6 brane through one of the NS5 branes
creates a D4 brane stretched between the NS5 brane and the D6 brane which
can then be moved away to infinity.

Classically the gauge coupling is given by the distance between the NS5
branes in the 6 direction, $\Delta$:
\begin{equation}
g^{-2} = \Delta / g_s l_s
\end{equation}
where $g_s$ and $l_s$ are the IIA string coupling and length respectively.

\subsubsection{Corrections: Bending and instantons}

To understand quantum effects in the gauge theory we must consider the
consistency of the brane configuration more carefully. In IIA string theory
the end points of the D4 branes on the NS5 branes are singular. However an
approximate description is that the D4 branes exert a force on the NS5 branes
causing them to bend. Considering only the directions not in common to both
branes, this leads to a logarithmic bending of the NS5 branes into the 6
direction along the 45 plane. This logarithmic variation of the separation
of the NS5 branes is interpreted in the field theory as the logarithmic
running of the gauge coupling. In this way the shape of the branes
incorporates the 1-loop effect in the field theory.

In ${\cal N}=2$ field theory there are no higher loop effects.
However there are still
non-perturbative effects due to instantons. These instantons can be seen
directly in the brane picture. It is known that a D(p-4) brane within a Dp
brane satisfies the 4 dimensional YM instanton equations \cite{BranesWithin}.
So D0 branes are
instantons within D4 branes. To interpret these D0 branes as instantons in
the 0123 spacetime we should consider Euclidean D0 branes whose world-line is
in the 6 direction so that they are contained within the D4 branes between
the NS5 branes \cite{Barbon1, Barbon2, BrodieD0}.
So now the field theory objects and quantum
effects have been mapped to D-branes and their properties. The problem is now
to solve the
theory by including all these effects. This can be done by ``lifting'' the
IIA configuration to M-theory \cite{Witten}.

\subsubsection{The Solution: Lifting to M-theory}

The advantage of considering the above configuration of branes in M-theory is
that the D4 branes and NS5 branes are in fact the same object, the M5 brane.
The intersection of D4 and NS5 branes in IIA is singular but this is smoothed
out in M-theory and in fact it is possible to consider all the D4 branes and
NS5 branes as a single M5 brane with complicated world-volume. However the
conditions for preserving ${\cal N}=2$ supersymmetry restrict the embedding
of the
M5 brane world-volume and it is possible to find the function describing this
embedding explicitly. Clearly this must incorporate the field theory 1-loop
effects through the shape of the M5 brane.
However the non-perturbative instantons
are also automatically included since D0 branes are simply Kaluza-Klein
momentum modes of compactified M-theory. (Here we consider ``momentum'' with
the 6 direction being Euclidean time, i.e.\ variation in $x^{11}$ wrt.\ 
$x^6$.) The embedding is in fact precisely given by the Seiberg-Witten curve
(times the trivial embedding of $R^4$ in the 0123 directions.)
The Seiberg-Witten differential can also be naturally identified in this
approach \cite{SWdiff}.
This relation between the IIA setup with separate classical, 1-loop
and non-perturbative effects, and the M-theory setup incorporating all effects
is shown in the following figure.

\vspace{1cm}
\fig{Solving HW setup via M-theory}
{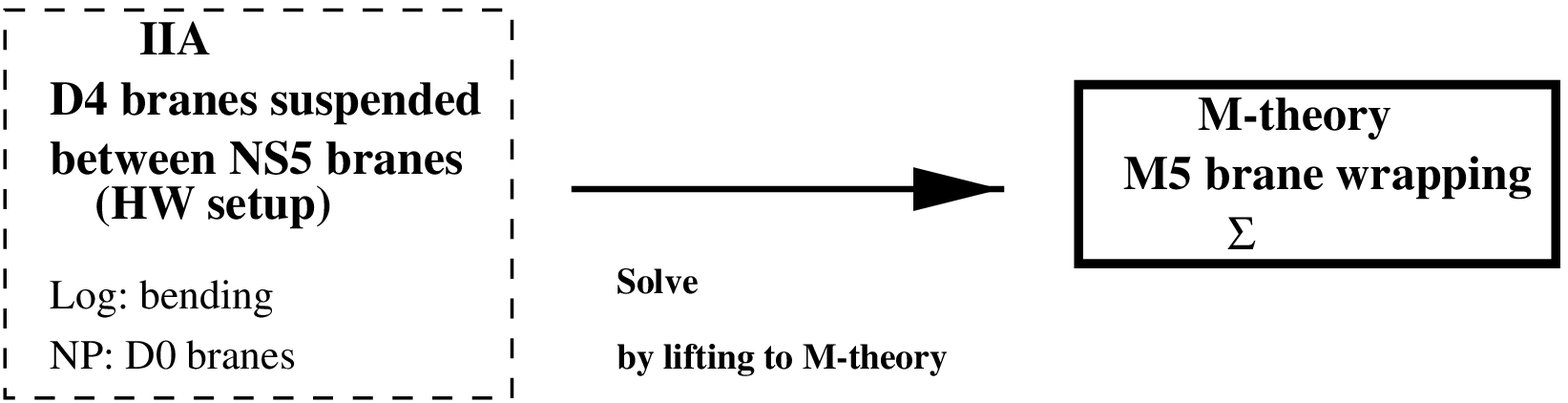}{9truecm}
\figlabel{\hw}

\subsubsection{Other Gauge Theories}

There are several ways to construct generalisations of the construction given
above. Of relevance to us are product gauge groups. These are easily described
by adding more NS5 branes. The number of D4 branes between 2 consecutive
NS5 branes determines an $SU(N)$ factor of the product gauge group. D6 branes
give matter in the fundamental representation of one of the $SU(N)$ factors
depending on which NS5 branes they are between. There is also matter in
the bi-fundamental of ``neighboring'' $SU(N)$ factors, coming from strings
within each NS5
brane stretching between D4 branes ending from either side.

The 6 direction can be compactified and this leads to ``elliptic'' models.
It is these models which we are particularly interested in here since we
will show that a T-duality can be performed on the 6 direction. From the
field theory viewpoint these are just product gauge groups as described
above where there is in addition bi-fundamental matter transforming under the
``first'' and ``last'' $SU(N)$ factors. \footnote{There is an additional $U(1)$
factor which is not frozen out in the brane picture in this case \cite{Witten}.
However, it decouples since all fields are neutral.}

${\cal N}=1$ gauge theories can also be constructed with these gauge groups.
This
is done by considering some of the NS5 branes to have world-volume 012345
and others to have world-volume 012389 \cite{EGK, rotate, brodie, EGK2,
Ooguri, QCD}.
Despite the reduction of the number
of supersymmetries it is still possible to understand many features of the
field theory from the branes. This includes a simple interpretation of
Seiberg duality \cite{seiberg} as an interchange of the NS5 branes
\cite{EGK,rotate}.

Gauge groups $SO$ and $USp$ can also be studied by including appropriate
orientifolds in the brane construction. In this paper we will only consider
SU gauge groups for simplicity. However,
our analysis should also extend to these cases.

\subsection{Branes as Probes}

A lot has been learned by putting Dp branes in a nontrivial
gravitational background and studying the p+1 dimensional field theory 
on the brane as a probe of the background geometry. We will focus
here on the approach of studying branes in an orbifold background
\cite{DouglasMoore, ken}. A variant of this approach has been chosen
by \cite{Probes,Seiberg3,Seiberg5,Seiberg6,Sen} where the probe moves 
in the background of a higher dimensional brane. Some of
these cases can be solved exactly by going to F-theory.

\subsubsection{D-branes at orbifolds}

We want to consider $N$ Dp branes on top of an orbifold singularity,
where the orbifold acts in the transverse spacetime. The formalism
for this procedure was developed in \cite{DouglasMoore,GP}.
In the following we will discuss the example of D3 branes at an orbifold,
since this leads to 4d physics. It has received a lot
of attention recently, since it has a dual description in
terms of supergravity on an AdS space \cite{Maldacena}.
Many exciting proposals have been made about this system, e.g.
it is conjectured to lead to theories that are finite \cite{KachruFinite,
Lawrence} even without
supersymmetry (at least for large $N$) and possess certain
S-duality properties.
The discussion generalizes in the obvious way to other Dp branes.

Hence consider $N$ D3 branes living in the 0123 space at an orbifold fixed
point,
where the orbifold group $\Gamma$ acts on the ``internal'' 456789 space.
The world-volume theory in the unorbifolded case is ${\cal N}=4$
$SU(N)$ SYM in 4d.
$\Gamma$ has to be a discrete subgroup of the $SO(6)$ acting on the internal
space. In the ${\cal N}=4$ setup this $SO(6)$ is the $SU(4)$ R symmetry.
Generically all SUSY is broken.
If $\Gamma \subset SU(3)$ ${\cal N}=1$ SUSY is preserved and for $\Gamma \subset
SU(2)$ ${\cal N}=2$ is preserved.

The ${\cal N}=4$ theory has the following matter content:
\begin{itemize}
\item the $U(N)$ vector
\item 4 adjoint gauginos
\item 6 adjoint scalars
\end{itemize}
where the scalars and fermions transform under the $SO(6)$ internal (R).

Besides the obvious action of $\Gamma$ on the scalars and fermions it also
acts on the vectors. The $U(N)$ gauge group is generated by the Chan Paton
factors living at the ends of the open strings connecting the D3 branes.
We therefore have to specify the action of $\Gamma$ on the Chan Paton factors.
If we want to describe D3 branes which are free to move away from
the orbifold fixed point we should restrict ourselves to embed
the orbifold group into the Chan Paton factors via
the regular representation $R$ of $\Gamma$, that is the $| \Gamma |$ dimensional
representation. Note that this representation is reducible and decomposes
in terms of the irreducible representations $r_i$ as $R=\oplus_i dim(r_i) r_i$.
Doing so we take into account the D3 branes and all its mirrors, as
is required if we want to have branes that are free to move away from the
fixed point (and than no longer coincide with their mirrors).
Other representations can be considered as well (at least at a classical
level) and lead to fractional branes \cite{DouglasMoore,Enhanced,
Diaconescu,C3orbi}. We will have more to say about these in what follows.
For the moment let's restrict ourselves to the regular representation.

Even though the orbifolded space is singular, string theory physics on the
orbifolded space is smooth. String theory automatically provides
additional modes in the twisted sectors that resolve the singularity.
In the case of an orbifold these twisted sectors are taken care
of by including also the $|\Gamma|$ mirror images of each original D3
brane.

So after all we start with an ${\cal N}=4$ $U(|\Gamma| N)$ SYM and just project
out degrees of freedom not invariant under $\Gamma$ where
$\Gamma$ acts:

\begin{itemize}
\item on the $SU(4)$ internal symmetry acting on the scalars
and fermions given in the obvious way by embedding of the geometric action
in internal spacetime
\item on the $U( |\Gamma| N)$ vector index according to $N$ copies of
the $|\Gamma|$ dimensional regular representation.
\end{itemize}

With this we get the following actions on the fields ($i,j$
labelling the columns of length $N$ of vector
indices transforming under the same irreducible representation
$r_i$ of $\Gamma$, $a=1,\ldots,6$ a vector
and $\alpha=1,\ldots,4$ a spinor index of the internal $SO(6)$ symmetry):

{\bf Vectors:}
$$A^i_j \rightarrow r_i \times \bar{r}_j A^i_j$$
leaving a $\prod_i U(dim(r_i) N)$ gauge group, since
$r_i \times \bar{r}_j$ contains the identity only for $i=j$.

{\bf Scalars:}
$$\Phi^{a i}_j \rightarrow a^6_{ik} r_k \times \bar{r}_j \Phi^{a i}_j$$
where $a^6_{ik}$ denotes the Clebsch Gordan coefficient in the decomposition
of $r_i \times {\bf 6} = \oplus_k a^6_{ik} r_k$, where
where {\bf 6} denotes the 6 dimensional representation of the scalars under
the R symmetry. We hence obtain $a^6_{ik}$ scalars transforming as
bi-fundamentals under the $i$th and $k$th gauge group. For $i=k$
this is interpreted as an 
adjoint. From the definition of $a^6_{ik}$ we can read off:
$$dim(r_i) \cdot 6 = \sum_k a^6_{ik} dim(r_k)$$

{\bf Fermions:}
The surviving fermions can be determined in the same manner as the
surviving scalars, just replacing $a^6_{ik}$ with $a^4_{ik}$ and
$a^{\bar{4}}_{ik}$. If $\Gamma \subset SU(3)$
we get $6 \rightarrow 3 + \bar{3}$, $4 \rightarrow 3 +1$ and
$\bar{4} \rightarrow \bar{3} +1$. This tells us that for every scalar
we also have a left and a right handed fermion and in addition
we also have an adjoint left and right handed fermion in every gauge group,
verifying that the spectrum is indeed ${\cal N}=1$ supersymmetric.
All potentials are derived from the unique ${\cal N}=4$ superpotential.

{\bf Examples: }
Consider for example the case of a $Z_n$ orbifold group generated
by $\alpha=e^{2 \pi i/n}$ acting on
the 3 complex dimensional internal space as

\begin{eqnarray*}
z_1 & \rightarrow& \alpha^{a_1} z_1 \\
z_2 & \rightarrow& \alpha^{a_2} z_2 \\
z_3 & \rightarrow& \alpha^{a_3} z_3 \\
\end{eqnarray*}

For $a_1 + a_2 + a_3 =0$ (mod $n$) we have $Z_n \subset SU(3)$, for
$a_1=1$, $a_2=-1$ and $a_3=0$ $Z_n \subset SU(2)$. One can see easily that
these cases indeed yield ${\cal N}=1,2$ supersymmetric spectra.

$|Z_n|=n$ and the $n$ irreducible representations are all one dimensional
($Z_n$ is Abelian). The $i$th representation is given by representing
$\alpha$ as $\alpha^i$ where $i$ runs from 0 to $n-1$. According to
the general rules the gauge group is hence just $U(N)^n$. To determine
the scalars and fermions we have to read off $a^6_{ik}$ and $a^4_{ik}$ from
the geometric action.
To do this note that $r_i \times r_j = r_{i+j}$ and that the reducible
6 and 4 dimensional representations decompose according to the embedding
of $Z_n$ in the spacetime $SO(6)$ as:
\begin{eqnarray*}
6 &\rightarrow& r_{a_1} + r_{a_2} + r_{a_3} +
		r_{-a_1} + r_{-a_2} + r_{-a_3} \\
4 &\rightarrow& r_{(a_1+a_2+a_3)/2}+r_{(a_1-a_2-a_3)/2}+
r_{(-a_1+a_2-a_3)/2}+r_{(-a_1-a_2+a_3)/2} \\
\bar{4} &\rightarrow& r_{-(a_1+a_2+a_3)/2}+r_{-(a_1-a_2-a_3)/2}+
r_{-(-a_1+a_2-a_3)/2}+r_{-(-a_1-a_2+a_3)/2} \\
\end{eqnarray*}
All the sums should be understood as mod $n$. The half-integer numbers
are due to the fact that to introduce spinors we should really go to the double
cover of $SO(6)$, namely $SU(4)$. This additional $Z_2$ action may not
commute with the $Z_n$. We see that the matter content consists of scalars
which are bi-fundamentals under the $i$th and the $(i+a_{\mu})$th gauge
group ($\mu=1,2,3$) and similarly for the fermions.

For $a_1+a_2+a_3=0$ (mod $n$) we see that we get one adjoint fermion
and one for every scalar ($(a_1 - a_2 - a_3)/2 =a_1$ and so on) yielding
the ${\cal N}=1$ spectrum. For the ${\cal N}=2$ choice we get an
adjoint multiplet and
the other chiral multiplets can be paired up into hyper multiplets,
as required. The ${\cal N}=2$ superpotential is provided automatically, since
all interactions are derived from the unique ${\cal N}=4$ potential.

\subsubsection{Moving onto the Coulomb branch: Fractional Branes}

So far we have identified the gauge theory of the $N$ D3 branes
living at the orbifold. Their Higgs Branch
can be easily identified as moving the branes away from the singularity,
breaking the $U(N)^k$ \footnote{We will now generally use the notation $k$
for the $Z_k$ orbifold action and $N$ for the number of D-branes.}
gauge group to its diagonal $U(N)$ subgroup.
This is just the gauge theory on the $N$ D3 branes. At energies
below the Higgs scale (given by the mass of open strings stretching
between the D3 branes and its distant images), we effectively
see the full ${\cal N}=4$ supersymmetry. The metric on the Higgs branch
is just the metric on the singular orbifold space.
If we turn on FI terms in the gauge theory, the Higgs branch
becomes a smooth ALE space. Turning on these FI terms therefore
corresponds in spacetime to blowing up the orbifold singularities.

But note that for vanishing FI terms this theory also has a Coulomb
branch, corresponding to turning on vevs for the adjoint scalars
in the vector multiplets, generically breaking each of
the $U(N)$ subgroups to its maximal torus.
How can we interpret this branch in spacetime?
The answer was given in \cite{Enhanced} (see also \cite{DouglasMoore}):
on the Coulomb branch, a single D3 brane splits into $|\Gamma|$ fractional
branes of equal mass $\frac{m_{D3}}{|\Gamma|}$. These fractional
branes can be interpreted as D5 branes wrapping the vanishing
cycles of the orbifold. Therefore they are stuck to the singularity
in the orbifolded part of spacetime, but are free to move in the
transverse space, giving rise to the Coulomb branch.
The example considered in \cite{Enhanced} was D0 branes moving
on an ALE space. In the orbifold limit these new degrees
of freedom supported on the Coulomb branch are precisely those that
are necessary for the Matrix description of enhanced gauge
symmetry for M-theory on the ALE space.

Looking just at a single fractional brane is achieved by
embedding the orbifold group into the Chan Paton factors via
a representation other than the regular one. This is consistent
with the fact, that only by choosing the regular representation
we can describe objects that are free to move away from the orbifold.
This identification can be proven by an explicit world-sheet calculation
\cite{DouglasMoore} showing that the fractional branes
carry charge under the RR fields coupling to D5 branes. This
charge vanishes if and only if one chooses the regular representation.
So instead of just obtaining fractional branes by splitting D3 branes
to move on to the Coulomb branch, one can just add
these fractional branes by hand. This is the approach chosen
in \cite{ALE1,ALE2} to explain the wrapped membranes in Matrix
theory. We will give a more detailed analysis of the
branches, parameters and the quantum behaviour of these theories as we
proceed.

\section{Duality between Hanany-Witten \\ and fractional branes}

\subsection{T-duality for fractional branes}

Consider again $N$ D3 branes at a transverse $A_{k-1}$ singularity, that
is we take the D3 branes to live in the 0123 space and put these
on top of an $A_{k-1}$ singularity which is represented by the following
$Z_k$ action
\begin{equation}
 z_1 = x_8 + i x_9 \rightarrow \alpha z_1, \quad z_2= x_6 + i x_7 
\rightarrow \alpha^{-1} z_2
\end{equation}
where $\alpha=e^{2 \pi i/k}$.
This $Z_k$ is a subgroup of the $SU(2)$ acting on $z_1$, $z_2$ and
we therefore have a gauge theory with ${\cal N}=2$ supersymmetry
in 4 dimensions. 
Since the D3 branes can move away from the singularity we should
choose to embed the orbifold group in the Chan Paton factors
via the regular representation, that is include the D3 brane and all its
$k-1$ images.
The corresponding gauge group is $U(N)^k$.
Upon T-duality in the 6 direction this can be mapped into an
elliptic Hanany-Witten setup, that is $k$ NS5 branes living
in 012345 directions, with $N$ D4 branes living in 01236 space
suspended between every pair of consecutive NS5 branes on the circle.
Let us label
the NS5 branes with $i=0,\ldots,k-1$. This kind of duality is by now
well known and was for example used to realize various non-trivial
fixed points in 6 dimensions via HW setups \cite{bk2,hztwo}.

Now consider the following setup:
\vspace{1cm}
\fig{The Hanany-Witten setup: N D4 branes stretching between
two of the NS5 branes}
{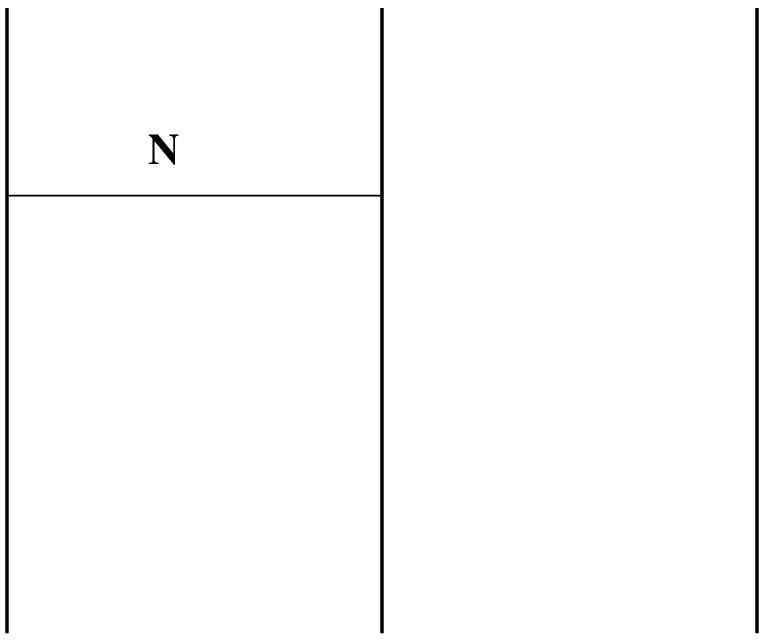}{7truecm}
\figlabel{\dual}

Figure (\dual) again displays a HW setup very similar to
the one above, this time describing just a single $U(N)$ gauge
group. We have just $N$ D4 branes stretching between the $i$th and
the $(i+1)$th NS5 brane.
It is natural to assume that the same T-duality
in the 6 direction as we used above now translates this into
$N$ fractional branes in the sense of \cite{Enhanced}, that
is $N$ D5 branes wrapping the vanishing homology two-cycle
$\sigma_i$ \footnote{where $\sigma_k= -\sum_i \sigma_i$ \cite{Enhanced}}
of the $A_{k-1}$ orbifold in the dual picture. Or again in the language of
\cite{Enhanced}, we consider the same orbifold group, but this
time choose to embed the orbifold group into the Chan Paton factors
via the 1d irreducible representation
$r_i$
associated with the $i$th node of the extended Dynkin diagram.
Or more generally, if we have $N_i$ D4 branes connecting the $i$th
NS5 brane with the $(i+1)$th NS5 brane the dual setup will again be given
by a $Z_k$ orbifold, this time with the orbifold group embedded in
the Chan Paton indices via a general representation $R$ given by
\begin{equation}
R= \oplus_i N_i r_i.
\end{equation}
As a first check note that the resulting gauge groups and matter contents
agree: both constructions yield a $\prod_i U(N_i)$ gauge
group with hypermultiplets transforming as bi-fundamentals
under neighboring gauge groups.
One may wonder whether this is consistent as a quantum
theory. Since, as we discussed above, the fractional 
branes are charged under the RR gauge fields we need `enough' non-compact
space in the transverse dimensions for this to be consistent. We
will have more to say about this when we discuss the quantum theory.
Similarly one can T-dualize any Dp brane stretching in between
NS5 branes as they appear in generalizations of the HW setup
to various dimensions to fractional D(p-1) branes
on an orbifold in the same manner.

\subsection{Supporting evidence for the proposed duality}

Since the precise metric describing a D4 brane ending on two 5 branes
is not known to us we cannot verify our proposal by a direct
computation.
However there is compelling evidence that it is indeed true.

\subsubsection{The 5 brane point of view}

Actually a very similar duality has already been proposed in
\cite{BrodieD0,Cherkis}. In both cases it was conjectured that
a brane ending on another brane is dual to some
object carrying fractional charge by considering the
world-volume theory of the brane on which the other brane ends,
that is in our case the NS5 brane. Let us briefly review
their arguments since they automatically give
some evidence for our case, too.

Let us first review the case discussed in \cite{BrodieD0}.
One starts with Euclidean D1 branes living in 36 space ending on
D3 branes which live in 0126 space. The D1 can end on the D3 in the
3 direction. This configuration describes the instanton corrections
in a HW realization of 3 dimensional gauge theories (after we suspend
the D3 branes in between NS5 branes in the 6 direction in the usual
fashion). One is interested in how the configuration transforms under
T-duality in the 3 direction. If the D1 branes stretch around the circle
of radius $R_3$, they simply T-dualize into Euclidean D0 branes
stretching along the 6 direction, which are known to yield
the instanton corrections to the corresponding gauge theory \cite{Barbon1,
Barbon2}. Since the D3 branes T-dualize into D4 branes this is now a
4d HW configuration. However the D1 branes can split on the D3
branes just like the D4 branes can split on the NS5 branes
in the setup we are interested in. The question of how the
system of Euclidean D1 branes ending on the D3 branes behaves under T-duality
was answered in \cite{BrodieD0} with the help of considerations
in the world-volume theory of the D4 brane (which is 4d since the D4 brane
itself should be thought of as being embedded in a HW setup).

The Euclidean D0 appears in this setup as a usual YM instanton. However
it is well known from field theory \cite{Kovner} that the gaugino
condensate in the corresponding YM
\begin{equation}
< \lambda \lambda > = \Lambda^3 = e^{- 8 \pi^2/N g_{YM}^2}
\end{equation}
is not produced by the instantons themselves but by fractional
instantons, that is configurations of
the form
\begin{equation}
A_{\mu}=\frac{1}{N} g \partial_{\mu} g^{-1}.
\end{equation}
From the field theory point of view, a usual instanton can shrink to
zero size and then split into $N$ fractional instantons. The $4 N \nu$
dimensional moduli space of $\nu$ instantons in 4d is then
(after fractionation of all the instantons) just given
by the positions of the $N \nu$ fractional instantons.
These objects are characterized by the properties that they carry
$\frac{1}{N}$th of the mass and charge of a regular instanton. Note
that this is precisely the property we are looking for to make contact
with our proposal: a Dp brane stretching around only $\frac{1}{k}$th of
the compact circle  T-dualizes into an object that has $1/k$ times
the mass and charge of the dual D(p-1) brane: a fractional brane.

This discussion does not directly apply to our setup, since
we don't consider Dp branes ending on other D-branes but in our
setup they end on NS5 branes. Under T-duality the NS5 branes turn into
a geometric background singularity and it is hard to discuss
their world-volume point of view. If we however consider a type IIB
HW setup (for example the original setup of \cite{HW} realizing
3d gauge theories), we can perform an S-duality before we T-dualize,
mapping the NS5 branes to D5 branes. This point of view
was taken in \cite{Cherkis}.

\subsubsection{Deformations and Moduli}

It is very nice to see how the various parameters and moduli
of the theory are realized as brane motions in the two
pictures and how they are mapped to each other.
First consider the case of all $N_i=N$, where the duality
is well established. This theory has two branches, a Coulomb
branch and a Higgs branch. As free parameters we can add FI terms $\xi_i$,
for each gauge group factor
which are triplets under the $SU(2)_R$ symmetry
and totally lift the Coulomb branch.
Since we have $k$ gauge group factors we have in addition
k gauge coupling constants $g_i^2$, which at least in the classical theory
are additional free parameters. Together with
the $\theta$-angles they form the complex coupling constants
$\tau_i=\frac{\theta_i}{2 \pi} + \frac{4 \pi i}{g_i^2}$

On the orbifold side the Higgs branch just corresponds to moving
the D3 branes away from the orbifold point in the ALE space. 
Therefore the Higgs branch metric coincides
with the metric of the transverse ALE space.
Since all the $N_i$ are equal in the HW setup all the D4 brane
pieces can connect to form $N$ full D4 branes which can then move
away from the NS5 branes in the 789 direction. As usual, the 4th real
dimension of this branch is given by the $A_6$ component of the
gauge field on the D3 branes.

When the D3 branes meet the NS5 branes in 789 space, that is at the origin
of the Higgs branch, they can separate into the pieces which are then
free to move around 45 space along the NS5 branes giving rise to the
Coulomb branch. In the orbifold picture the same process is
described by the $N$ D3 branes splitting up into $k N$ fractional
branes. These are now localized at the singularity in 6789 space,
but are also free to move in the 45 space.

Now consider turning on the FI terms $\xi_i$. For the
orbifold this
corresponds to resolving the singularity by blowing up the $i$th vanishing
sphere $\sigma_i$. The mass of the states on the Coulomb branch
(which are after all D5 branes wrapping these spheres) become
of order $\xi_i^2$. The Coulomb branch is removed as a supersymmetric
vacuum. In the dual picture the FI terms correspond to motion of the
NS5 branes in 789 space. This resolves the singularity
since the 5 branes no longer coincide and again the mass of states on
the Coulomb branch (which in this case are D4 brane stretching
between the displaced NS5 branes) is of order $\xi_i^2$.

Last but not least we have to discuss the coupling constants. On the HW side
they are simply given in terms of the distances $\Delta_i$
between the NS5 branes in the 6 direction, or to be more precise
\begin{equation}
g_i^{-2}=\Delta_i / g^A_s l_s
\end{equation}
where $g^A_s$ and $l_s$ denote string coupling constant and string length
of the underlying type IIA string theory
respectively. Since we are dealing with a theory with compact 6 direction
we have
\begin{equation}
\sum_i \Delta_i = R_6
\end{equation}
where $R_6$ is the radius of the 6 direction.
The easiest case to consider is if all $\Delta_i=\Delta$, that is
the NS5 branes are equally spaced around the circle.
In this case we get $g_i^{-2}=g^{-2}=R_6 / g^A_s k l_s$. Applying T-duality
to type IIB this
means that $g_i^{-2}=1 / g^B_s k$ since $g^B_s=g^A_s l_s/R_6$, in accordance
with the orbifold analysis \cite{DouglasMoore,Lawrence}.
According to our duality hypothesis the same configuration can be
also viewed as $N_i$ fractional branes, that is $N_i$ D5 branes
wrapping the $i$th vanishing cycle. According to \cite{Lawrence}
in this case the coupling constant is given in terms of fluxes
of two-form charge on the vanishing sphere
\begin{equation}
\tau_i=\int_{\sigma_i} B^{RR} + i \int_{\sigma_i} B^{NS}
\end{equation}
At the orbifold
the values of the B-fields are fixed to yield again $g_i^{-2}=1/g^B_sk$
\cite{Enhanced}.
Since we are however dealing with D5 branes wrapping the vanishing
sphere, $B^{NS}$ always appears in the combination ${\cal F}=F-B^{NS}$,
where F denotes the field strength of the gauge field on the
D5 brane. We therefore expect that the imaginary part of $\tau_i$, that
is $\frac{4 \pi}{g_i^2}$, is actually given by $\int_{\sigma_i} {\cal F}$.
The full gauge coupling is hence given by a constant piece from
the fixed value of the $B^{NS}$ field at the orbifold point
plus some modifications due to Wilson lines. Under
T-duality these Wilson lines translate into the positions
of the NS5 branes on the circle, as expected.

It is easy to see that this discussion generalizes without any problems
to arbitrary values of $N_i$. In addition we can introduce D6
branes in the HW setup living in 0123789 space to give extra matter. These
simply T-dualize into D7 branes wrapping the ALE space in the dual picture.

\subsubsection{Quantum Behaviour}

All considerations so far have been purely classical. Now let us
discuss the quantum behaviour. Since we are dealing with ${\cal N}=2$
SUSY there are only two kinds of corrections, one-loop and
non-perturbative corrections. All higher loop contributions
vanish exactly. The one loop beta function gives rise to
the usual logarithmic running of the coupling constant. The
instanton contributions have been summed up by the Seiberg-Witten solution.

On the HW side it is well known how to incorporate quantum corrections.
In the full string theory the D4 brane ending on the NS5 brane
actually has a back-reaction on the NS5 brane leading to a logarithmic
bending of the NS5 brane. Since the coupling constant is given
in terms of distance between the NS5 branes this fact just
reflects the 1-loop correction, the running of the gauge coupling.
In addition this bending actually freezes out all the $U(1)$ 
factors yielding just a $\Pi_i SU(N_i)$ gauge group.
The FI terms no longer exist. Motions
in the 789 space are now associated to baryonic expectation values.

The nonperturbative
effects in this case are given by Euclidean D0 branes stretching
along the 6 direction \cite{Barbon1,Barbon2,BrodieD0} 
using the well known relation that
SYM instantons inside a Dp brane are represented by D(p-4) branes
\cite{BranesWithin}.
Like the D4 branes themselves these Euclidean D0 branes can of
course split on the NS5 branes and become fractional D0 branes 
\cite{BrodieD0}. According to the same T duality in
the 6 direction we applied to the D4 branes they should become
fractional D(-1) branes, that is Euclidean D1 branes wrapping
the vanishing cycles. This fits nicely with the picture
that this way the instanton corrections on the orbifold
side are again given by D(p-4) branes living inside the
Dp brane.

On the HW side the instanton corrections can actually be summed up.
The D0 branes are just momentum modes along the hidden 11 direction.
They are incorporated by lifting the configuration to M-theory 
\cite{Witten}. Similarly the instanton contributions on the orbifold
side can also be summed up using the tools of geometric
engineering. We will have more to say about this connection in 
the next section.

But how does the 1-loop effect arise in the orbifold picture?
Note that the fractional branes carry charge under the appropriate
RR fields, as can be born out by a world-sheet computation
of the corresponding tadpoles
\cite{DouglasMoore}.
They represent a charge sitting in the space transverse to
the D3 brane and transverse to the orbifold. In our example
this is the 2 dimensional 45 space. In two dimensions the Green's functions
are given by logarithms. Therefore we would expect a theory of
a charged object in 2 dimensions to be plagued by divergences.
To be more precise we have the following trouble: we want
to consider effectively a 3 brane sitting in 6d spacetime.
Since the transverse space is only 2d, the corresponding
4-form vector potential $C^{(4)}$ grows logarithmically.

From the D5 brane point of view the 3 brane vector potential
couples to the world-volume theory via
\begin{equation}
 \int dx^6 {\cal F} \wedge
C^{(4)}.
\end{equation}
After compactification on $\sigma_i$ this leads to a term in the
3 brane action of the form
\begin{equation}
\int_{\sigma_i} {\cal F} \cdot \int dx^4 C^{(4)}.
\end{equation}
But recall that we identified $\int_{\sigma_i} {\cal F}$ as the
gauge coupling $\frac{4 \pi}{g_i^2}$.
The growing of $C^{(4)}$ can be absorbed by introducing an effective
running coupling constant.
The divergences we encounter are just due to the 1-loop running
of the gauge coupling! Charge neutrality,
which is obtained if we choose the regular representation,
corresponds
to finiteness \cite{DouglasMoore,Lawrence}. In this way we can see that
the correct dependence of the beta function on $N_c$ and $N_f$ is reproduced.
This can most easily be seen by observing that the charge clearly depends
linearly on the number of fractional branes and is zero when $N_f = 2N_c$,
corresponding to the case where the fractional branes can all combine to
form a full brane. The relative contributions originate from the
(self-)intersection numbers of the spheres $\sigma_i$ given by the
(extended) Cartan matrix (here for $A_{k-1}$).

In this way the bending of the
branes directly translates into the fall-off of the RR field in transverse
space in the orbifold picture. The correspondence also
holds in other dimensions (for D5  branes
the transverse space is 0 dimensional and we need neutrality
reflecting anomaly freedom of the underlying gauge theory, in 5,4,3,2,...
dimensions we get linear, logarithmic, $1/r$, $1/r^2$, ... fall off
for the field strength). This reflects the appropriate running of the gauge
coupling just as in the HW picture.
The following diagram illustrates this duality.

\vspace{1cm}
\fig{Duality relation between HW setup and fractional branes}
{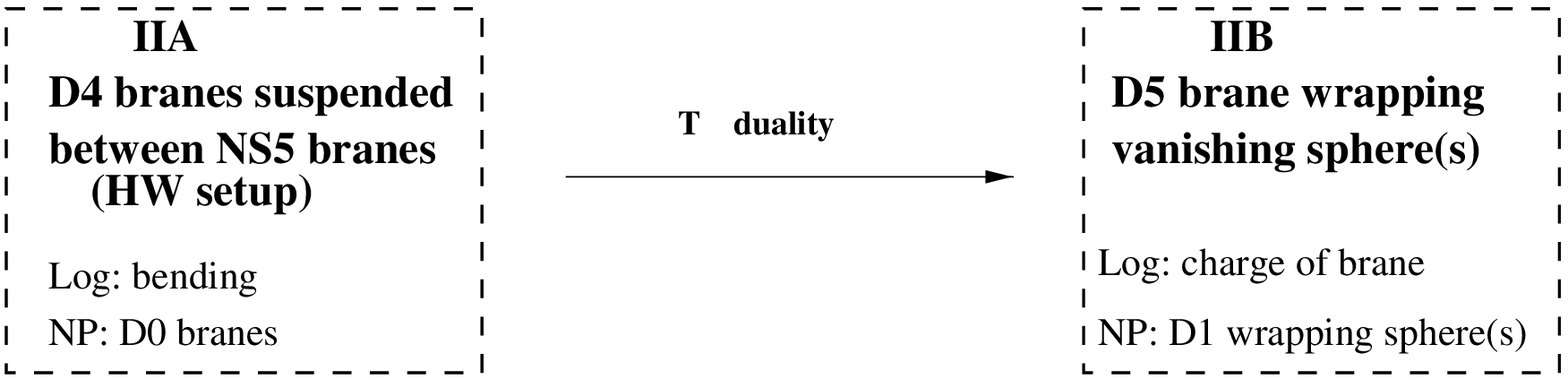}{10truecm}
\figlabel{\tdual}

\section{Connecting the pieces}

The duality we proposed above is the missing piece to connect the various
approaches to embedding gauge theories into string theory, making
their quantum behaviour manifest. We described these approaches above. Here
we will show how they are related.

\vspace{1cm}
\fig{Connection between the various approaches. The non-compact
Calabi Yau spaces are ALE spaces fibered over a sphere (or several
intersecting spheres). These can be T-dualized via T-duality
on the fiber. The last T-duality, connecting to
the HW picture is the one discussed in this work.
}
{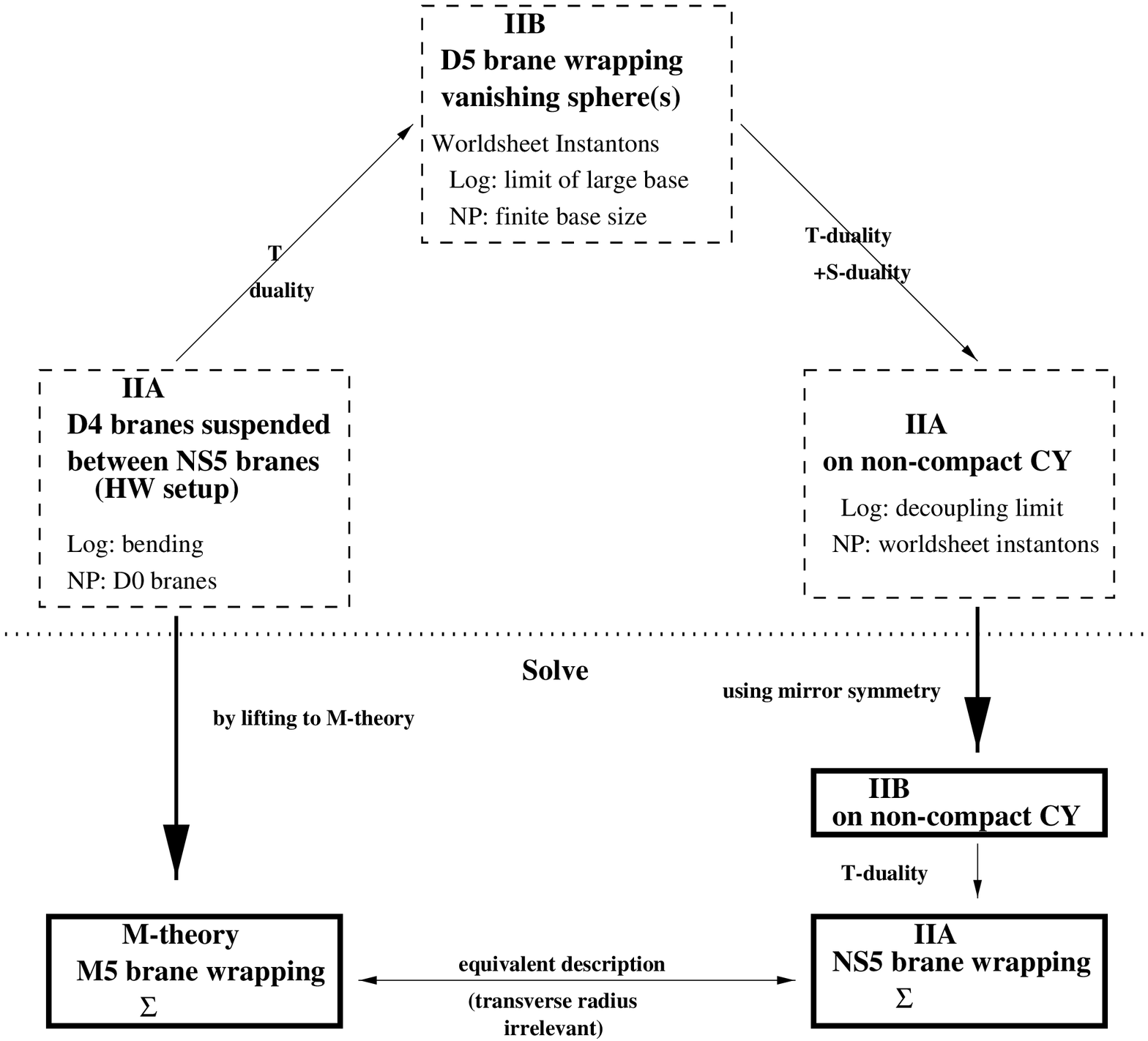}{14truecm}
\figlabel{\dia}

In Figure (\dia) the connections are displayed diagrammatically, putting
together the pieces of Figures (\eng), (\hw) and (\tdual).
For the
geometric engineering 
approach of \cite{EngineerII} the starting point is IIA
string theory on a non-compact Calabi-Yau which is an ALE space
fibered over some base (a non-compact version of a K3 fibration).
Quantum corrections can be solved by using mirror symmetry
to a IIB picture, as reviewed above. A T-duality on the ALE
fiber can be used to map this to a IIA NS5 brane wrapping
the Seiberg-Witten curve \cite{Klemm, EngineerII}. While in this approach
the solution is obtained via mirror symmetry, this last
T-duality leads to the very nice interpretation
of the SW curve: it appears as the physical object the
5 brane is wrapping.

Applying the same T-duality, which we have just applied to the IIB solution,
directly on the ALE fiber in the
IIA setup (which is the starting point for the Geometric Engineering approach),
followed by an S-duality in the resulting IIB string theory,
it is straight-forward to connect
this to a setup of D5 branes wrapping the base of the fibration.
For the geometric engineering approach to ${\cal N}=1$ theories
\cite{EngineerI,EngineerDual} this picture of branes wrapping
the base of an ALE fibration is actually the starting point,
as we will discuss in the next section. 
This way we can make direct contact to the language of fractional
branes \cite{DouglasMoore,Enhanced,Diaconescu} where
one discusses the world-volume theory of branes wrapping various
cycles.

Using the T-duality proposed in this paper one can translate this
into the very intuitive language of a HW setup \cite{HW},
which finally can be solved for by lifting to M-theory \cite{Witten}.
This way we not only obtain the same solution (the SW curve), but
also the same physical interpretation: the SW curve appears directly
as the space an M5 brane wraps on 
\footnote{In the spirit of \cite{SeibergSethi} the difference between
IIA and M5 brane is irrelevant, since the radius of the transverse
circle does not affect the low-energy SYM on the brane.}.

In the following we will discuss the steps in the above chain of dualities
in more detail. There are two important points one should account
for. For one thing there are the quantum corrections. Since we are
dealing with ${\cal N}=2$ there are only two types of these corrections:
the 1-loop contribution which gives rise to the logarithmic
running of the gauge coupling and the instanton corrections.
It is interesting to see how these quantum effects are incorporated
in the various pictures and how they are finally solved for.

The other thing one should treat with some care is how the singularity
type affects the gauge group.
In all the examples we are considering there are always two
pieces of information, which we will refer to as the singularity
types determining the ``gauge group'' and the ``product structure''.
This is easiest to understand in a particular example.
Consider $N$ D3 branes at an $A_{k-1}$ singularity. This gives rise
to an $SU(N)^{k}$ gauge group. In this example $N$ determines
the ``gauge group'' (the size and type of the single gauge group factors)
and $k$ the ``product structure'' (we have $k$ $SU(N)$ factors). Since
in the various duality transformations we are using we repeatedly
map branes into singularities and vice versa it is quite important
to distinguish which ingredient in the picture determines
gauge group and product structure. 
In the cases where the gauge group is determined by a geometric
singularity generalization to D or E type singularities leads
to $SO$ or $E_{6,7,8}$ gauge groups, while in the
cases where the product structure is determined by a geometric
singularity the generalization to D or E type just leads to a more
involved product of $SU(N)$ groups determined by
the corresponding extended Dynkin diagram. For example in the D3
brane case $N$ D3 branes at a $E_6$ singularity lead to
an $SU(N)^3 \times SU(2N)^3 \times SU(3N)$ gauge group.

We will
start with the HW setup, since this seems to be the most
intuitive approach. 

{\bf HW $\rightarrow$ fractional branes: }
This is basically the duality we have discussed in this work. A system
of $k$ NS5 branes with $N_i$ D4 branes suspended between the
$i$th and the $(i+1)$th NS5 brane is dual to an
$A_{k-1}$ orbifold singularity with $N_i$ fractional branes associated
to the $i$th shrinking cycle. While in the HW setup the gauge
group is determined by the D4 branes and the NS5 branes determine the product
structure, the corresponding roles are played by D3 branes and the orbifold
type in the dual picture. That is, the vanishing 2-cycles are given by
spheres intersecting according to the
extended Dynkin diagram. The gauge group can be generalized
to D type by including orientifold planes on top of the D-branes.
It is not known how to achieve E type gauge groups this way. Using an E type
orbifold in the fractional branes only affects the product structure.

The 1-loop quantum effects correspond to bending of the NS5 branes in HW
language and get mapped to the logarithmic Green's functions in the
orbifold picture, as discussed in the previous section. Non-perturbative
effects are due to Euclidean D0 branes, which are mapped by the same T-duality
into Euclidean D1 branes wrapping the vanishing 2-cycles (fractional
D-instantons).

{\bf Fractional branes $\rightarrow$ IIA on non-compact CY: }
We can now apply a different T-duality on the setup described by
the fractional branes. First we S-dualize, taking the $N_i$ D5 
branes into NS5 branes
and the Euclidean D1 branes into fundamental strings (that is world-sheet
instantons). Performing a T-duality in the overall transverse space
(that is 45 space) the NS5 branes turn into an $A_{N_i}$ singularity
according to the duality of \cite{dualNS} which we have already used
several times. The vanishing 2 cycle, which is still
the space built out of the $k$ spheres intersecting according
to the extended Dynkin diagram, stays unchanged.
The fact that the NS5 branes only wrap parts of this base ($N_i$ NS5 branes
on the $i$th sphere $\sigma_i$) translates into the fact that
the type of the ALE fibers over the base changes from one sphere to the other.
From the IIA point of view this looks like T-duality
acting on the fibers as described in
\cite{Klemm}. Since now everything is geometric, generalizations to E type
are straight forward: the product structure is determined by
the intersection pattern of the $k$ spheres, the ADE type of the
fiber determines the gauge group. This way we can even
engineer products of exceptional gauge groups. 

The non-perturbative effects are now due to world-sheet instantons,
as expected. The log-corrections coming from one loop are incorporated
in the particular limit one has to choose to decouple gravity 
\cite{DecouplingLimit}. The system can be solved via local mirror
symmetry.

\section{Generalizations to ${\cal N}=1$}

All the approaches discussed in the previous section have been generalized
to capture ${\cal N}=1$ supersymmetric gauge dynamics. Even though different
approaches are more or less suited for different questions, the above
pattern of connections between the various approaches generalizes
to this case as well. As a starting point let us take the HW setup,
since this case was discussed extensively in the literature \cite{EGK,
brodie,EGK2}. The breaking of the supersymmetry is achieved
by rotating some of the NS5 branes into NS5' branes \cite{rotate}, which now
live in 012389 space. The theory can still be solved via a lift to M-theory
\cite{Ooguri,QCD}.

The first step in the chain of dualities is to translate this into
fractional D3 branes sitting at an orbifold singularity. According
to the standard lore the system of $k$ NS5 branes and $k'$ NS5$'$ branes
T-dualizes into a $Z_k \times Z_{k'}$
orbifold action on the 456789 space transverse to
the brane given by the combination of the group actions
\begin{equation}
 z_1 = x_8 + i x_9 \rightarrow \alpha z_1, \quad z_2= x_6 + i x_7
\rightarrow \alpha^{-1} z_2
\end{equation}
\begin{equation}
 z_2 = x_6 + i x_7 \rightarrow \alpha' z_1, \quad z_3= x_4 + i x_5
\rightarrow {\alpha'}^{-1} z_3
\end{equation}
This is a particular case of an orbifold group $\Gamma \subset SU(3)$
discussed in Section 2. Again D4 branes stretching all around the
circle become D3 branes, while D4 branes ending on the NS5 branes
become fractional D3 branes. In this case the fractional D3 branes
are D7 branes wrapping a vanishing 4 cycle S, which is given by a
fibration of $P_1$ over $P_1$ (a Hirzebruch surface), where
the $P_1$s are just the vanishing spheres associated with any one
of the two orbifold actions. 
Instanton corrections are given by fractional D-instantons, that
is Euclidean D3 branes wrapping the vanishing 4 cycle. 

While the NS5 branes still bend, this bending no longer encodes directly
the $\beta$ function. However a different T-duality can be applied to
bring the theory into a slightly modified form recently
considered by \cite{HanZaff}, containing
D5 branes suspended in rectangles spanned by two types
of NS5 branes.\footnote{One applies 2 T-dualities in the 4 and the 7 direction
instead of one in the 6 direction to the orbifold described above. This yields
the Hanany-Zaffaroni setup (up to some renaming of directions).}
In this picture the bending again carries information about
the $\beta$ function: in the case of finite theories there
is no bending \cite{HanFinite,Gremm}. In some special cases it is possible
to find the coefficient of the beta function in this setup \cite{ArmBrand}.
The orbifold picture
leads to a very similar conclusion: if we choose the
regular representation charge neutrality still implies finiteness
\cite{KachruFinite}.
For non-zero net charge in the orbifold space we encounter divergences
which should again be interpreted as being due to the running of the gauge
coupling.

Note that this is precisely the setup used as a starting point
in the Geometric Engineering approach to ${\cal N}=1$ theories
\cite{EngineerI,EngineerDual}. It can also be reformulated in terms
of F-theory, where now the position of the D7 branes is encoded
in the singularities of the elliptic fiber over the surface S,
which is the compact part of the base of the elliptic fibered CY 4 fold
that is used in F-theory compactifications yielding ${\cal N}=1$ in 4d.
Since now the 7 branes (which in our language from above determine
the gauge group and not the product structure of the gauge group)
are translated into geometric singularities, generalizations
of this setup to $SO$ and exceptional gauge groups are straight-forward.
It would be nice to solve these theories (like the ${\cal N}=2$ ones)
via a local mirror symmetry of the 4-fold (which can be thought of as
T-duality on a 4 cycle and hence takes IIB (F-theory) back to IIB (F-theory)).
Even though there has been some progress in this direction \cite{Lerche,
Mayr}, the solution of the problem remains for future work.

\section{Conclusions}

In this work we have presented a relation between various approaches
to obtain field theories as limits of string theory. The
last piece in the puzzle was a T-duality between HW setups
and fractional branes at orbifolds, which is a generalization
of the well known T-duality relating elliptic HW setups to
regular branes at orbifolds. 

A lot has been learned about latter ones using the recently discovered
correspondence of conformal field theory with supergravity
on AdS spaces \cite{Maldacena, KachruFinite}. 
It would be very interesting to see whether there is also a supergravity
description corresponding to the large $N$ limit of non-conformal gauge
theories realized in terms of fractional branes. This
space will no longer be AdS, since conformal symmetry is lost, but
it should encode the logarithmic divergences due to the
running of the gauge coupling. 

\section*{Acknowledgements}
We would like to thank Ilka Brunner for useful discussions and
Michael Douglas and Amihay Hanany for helpful correspondence. The research
of A.K. and D.S. is supported by Deutsche
Forschungsgemeinschaft.

\bibliography{connect}
\bibliographystyle{utphys}
\end{document}